\documentclass[12pt]{article}

\usepackage{epsfig}

\usepackage{amssymb}

\def\be{\begin{equation}}
\def\ee{\end{equation}}
\def\ben{\begin{displaymath}}
\def\een{\end{displaymath}}
\def\ba{\begin{array}{c}}
\def\ea{\end{array}}

\newcommand{\ed}{\end{document}}

\begin{document}

\titlepage
%\vspace*{2cm}

\begin{center}{\Large \bf

Schr\"{o}dinger equations with indefinite effective mass

%es $m(x)$ energy- and coordinate-dependent

 \vspace*{1.2cm}

}\end{center}

%\vspace{10mm}

\begin{center}

%\newpage
Miloslav Znojil  \vspace{3mm}

Nuclear Physics Institute of Academy of Sciences of the Czech
Republic, 250 68 \v{R}e\v{z}, Czech Republic,

\vspace{2mm} email: znojil @ ujf.cas.cz

\vspace{2mm} phone: 00420 266173286

\vspace{3mm}

and \vspace{3mm}

  G\'{e}za L\'{e}vai

\vspace{3mm}

Institute of Nuclear Research of the Hungarian Academy of Sciences
(ATOMKI),

PO Box 51, H-4001 Debrecen, Hungary

\vspace{2mm} email: levai @ namafia.atomki.hu

\vspace{3mm}

\end{center}

\vspace{5mm}

\vspace{5mm}
%\newpage

\section*{Abstract}

The consistency of the concept of quantum (quasi)particles
possessing effective mass which is both position- and
excitation-dependent is analyzed via simplified models. It is shown
that the system may be stable even when the effective mass
$m=m(x,E)$ itself acquires negative values in a limited range of
coordinates $x$ and energies $E$.

\newpage

\section{Introduction}

Non-relativistic quantum dynamics of point particles is most often
studied via ordinary differential Schr\"{o}dinger equation
 \be
 -\frac{\ \hbar^2}{2m}
  \frac{d^2}{d{x}^2}\, \Psi(x)
 +V(x) \, \Psi(x) = E \,\Psi(x)\,,
 \ \ \ \ \ \ \ \Psi(x) \in L^2(\mathbb{R})
 \label{SEor}
  \ee
where the real function $V(x)$ characterizes an external local
potential while the particle mass $m>0$ is just a given constant.
Various generalized forms of this equation were introduced due to
the practical needs of the description of motion of a particle or
quasiparticle inside a medium. The medium may make the mass
position-dependent \cite{coordep}. The phenomenological appeal of
the spatial variability of the effective mass is accompanied by some
additional formal merits of the generalization, say, in the context
of supersymmetric quantum mechanics \cite{Plastino}.

In the majority of papers which studied the models with $m=m(x)$ the
authors assumed that $m(x)>0$ \cite{two}. The more we were surprised
when we found \cite{Siegl} that in the so called ${\cal
PT}-$symmetric version of the quantum Coulomb problem the stability
of the system required an anomalous, {\em asymptotically negative}
choice of the coordinate-dependent effective mass. This observation
re-attracted our attention to quantum models with $m=m(x)$ and, in
particular, to their subset in which one encounters the anomalous
$m(x)<0$, in some nonempty interval of coordinates at least.

In what follows we intend to describe some results of our analysis.
Firstly, in section \ref{instab} we introduce an elementary solvable
Schr\"{o}dinger equation with a piecewise constant mass such that
$m=m(x)<0$ for $x \in (-1,1)$. We shall demonstrate that the
bound-state spectrum of energies of such a model is unbounded from
below so that the model cannot be interpreted as realistic due to
its immanent instability. This result supports our expectations that
for the models with indefinite mass $m(x)$ such an instability will
be generic.

One of the possible methods of elimination of similar pathologies is
then proposed in section \ref{stabil}. We recall the general
Feshbach's definition of the effective quantities in quantum theory
(cf. ref. \cite{Feshbach}) and conjecture that the apparent
anomalies in the behavior of the benchmark square-well model of
section \ref{instab} are not realistic. We argue that their
emergence should be attributed to our unfounded complete suppression
of the necessary variability of the effective mass with the energy.
On this basis we propose and describe our first exactly solvable
model with $m=m(x,E)$ in section \ref{stabil} and its alternative
version in section \ref{secondex}.

Our main result is that both of these amended benchmark models may
remain stable (and, hence, acceptable for phenomenological
purposes), provided only that the mass $m(x,E)$ stays merely
anomalous (i.e., negative) in a restricted range of its arguments
(i.e., in finite intervals of coordinates $x$ and energies $E$).

A few formal aspects and possible consequences of the reinstalled
mathematical consistency and acceptability of Schr\"{o}dinger
equations with sign-changing effective masses $m(x,E)$ will be
finally mentioned in section \ref{discussion}. We shall re-emphasize
there the phenomenological as well as purely theoretical appeal of
the use of effective masses $m(x,E)$ which change their sign.
Whenever it happens just locally in both $x$ and $E$, one may
encounter no contradiction with expectations and/or with the general
principles of quantum mechanics.

\section{The instability of quantum systems with a
locally negative effective mass $m=m(x)$ \label{instab}}

%\subsection{Toy model}

In the majority of applications one accepts the requirement of the
positivity of the mass $m(x) > 0$ as natural. In the light of
Ref.~\cite{Siegl}, such a rule might have its exceptions. We should
admit that in the latter paper the weird-looking asymptotic
negativity of the effective mass can in fact be attributed to our
rather technical postulate of the loss of the observability of the
coordinate $x$ in the asymptotic region \cite{negama}. {\em Vice
versa}, we believe that the effective mass $m(x)$ {\em should}
remain asymptotically positive whenever the asymptotic coordinates
$x$ remain real. In this context, our present letter may be read as
motivated by the question of possible existence of some physically
consistent scenarios admitting a {\em negative} effective mass
$m(x)<0$, say, in some not too large {\em finite interval} of
coordinate $x\in \mathbb{R}$.

For the sake of definiteness let us simplify the mathematics of such
a conceptual problem by considering just the toy model in which the
energy spectrum is discrete and in which one works with the deep
square-well potential
 \be
 V(x)=
 \left \{
 \begin{array}{cc}
 +\infty,&\ \ \ |x| > L > 1\,,\\
 0, \ \ \ & \ \ \ |x| < L \,.
 \end{array}
 \right .
 \label{potencal}
 \ee
Moreover, in our models the mass will be piecewise constant, say, as
follows,
 \be
 m(x)=
 \left \{
 \begin{array}{cc}
 1,&\ \ \ |x| \in (1,L)\,,\\
 m_0, \ \ \ & \ \ \ |x| < 1 \,.
 \end{array}
 \right .
 \label{pushy}
 \ee
Unfortunately,  whenever $m_0$ is negative, the spectrum of the
bound-state energies of such a model becomes unbounded from below.
This makes the whole system unstable with respect to perturbations
so that its physical meaning becomes highly questionable. Still, it
makes good sense to understand this counterintuitive and apparently
discouraging fact in a more quantitative detail.

%\subsection{A toy-model-based explanation of the instability \label{hury} }

%
%
%VVVVVVVVVVVVVVVVVVVV

Let us consider the toy-model Schr\"{o}dinger equation
 \be
 -\frac{\ \hbar^2}{2m(x)}
  \frac{d^2}{d{x}^2}\, \Psi(x)+V(x) \, \Psi(x) = E \,\Psi(x)\,
 \label{SEorex}
  \ee
with potential (\ref{potencal}) and mass (\ref{pushy}) where, say,
$m_0=-1$. Let us also restrict our attention, for the sake of
brevity, to the mere even-parity bound-state solutions $\Psi(x) =
\Psi(-x)$ such that $\Psi(-L) = \Psi(L)=0$.

In units $\hbar^2/2=1$ and in the first step of analysis we shall
assume that $E = k^2$ is non-negative. This means that we have to
solve two differential equations,
 \be
 \begin{array}{cc}
 -  \frac{d^2}{d{x}^2}\, \Psi(x) = k^2 \,\Psi(x)\,\ \ &
  \ \ x \in (-L,-1)
  \,,\\
  - \frac{d^2}{d{x}^2}\, \Psi(x) = - k^2 \,\Psi(x)\,\ \ &
  \ \ x \in (-1,0)
  \,,\
 \ea
 \ee
with the respective explicit solutions
 \be
 \begin{array}{cc}
 \Psi(x) = \sin k(x+L)\,\ \ &
  \ \ x \in (-L,-1)
  \,,\\
   \Psi(x) = D\,{\rm cosh}\,kx \ \ &
  \ \ x \in (-1,0)
  \,
 \ea
  \ee
which satisfy the usual physical boundary conditions. In addition,
these solutions must also satisfy the usual mathematical matching
conditions at $x=-1$,
 \be
 \ba
 D\,{\rm cosh}\,k=\sin k(L-1)\,,\\
 -D\,{\rm sinh}\,k=\cos k(L-1)\,.
 \ea
 \ee
This determines the eigenvalues $E_n=k^2_n$ via an elementary
transcendental equation
 \be
 {\rm tanh}\,k \times \tan k(L-1)= -1\,
 \label{prveq}
 \ee
(cf. the two samples of its graphical solution in
Fig.~\ref{obr1ag}).

\begin{figure}[h]                     %instead of \begin{figure}[t]
\begin{center}                         %instead of \begin{center}
\epsfig{file=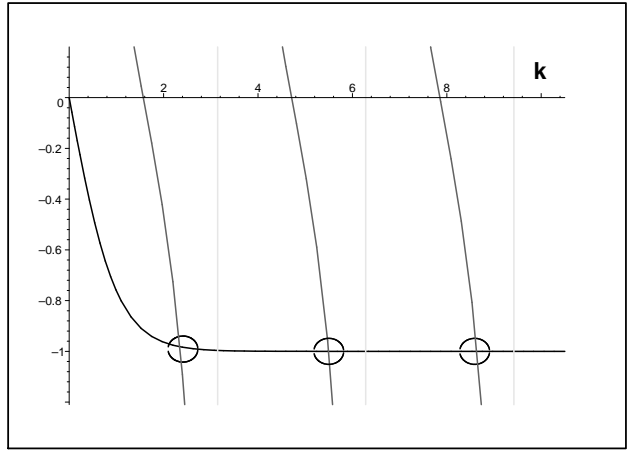,angle=270,width=0.6\textwidth}
\end{center}                         %instead of \end{center}
\vspace{-2mm} \caption{Graphical solution of Eq.~(\ref{prveq}) at
$L=2$ (the small circles mark the intersections of $-{\rm tanh}\,k$
with  $\cot k(L-1)$).
 \label{obr1ag}}
\end{figure}

In the same units $\hbar^2/2=1$ and in the second step let us set $E
= -\kappa^2$ and write down the two corresponding equations,
 \be
 \begin{array}{cc}
 -  \frac{d^2}{d{x}^2}\, \Psi(x) = - \kappa^2 \,\Psi(x)\,\ \ &
  \ \ x \in (-L,-1)
  \,,\\
  - \frac{d^2}{d{x}^2}\, \Psi(x) =  \kappa^2 \,\Psi(x)\,\ \ &
  \ \ x \in (-1,0)
  \,,\
 \ea
 \ee
with the respective explicit solutions
 \be
 \begin{array}{cc}
 \Psi(x) = {\rm sinh}\, \kappa(x+L)\,\ \ &
  \ \ x \in (-L,-1)
  \,,\\
   \Psi(x) = F\,\cos\,\kappa x \ \ &
  \ \ x \in (-1,0)
  \,.
 \ea
 \label{anonym}
 \ee
The matching conditions at $x=-1$ yield the two constraints,
 \be
 \ba
  {\rm sinh}\, \kappa (L-1)= F\,\cos \kappa\,,\\
 {\rm cosh}\, \kappa (L-1)= F\,\sin \kappa\,.
 \ea
 \ee
The resulting implicit definition of the negative-energy eigenvalues
$E_n=-\kappa^2_n$ is then obtained,
 \be
 \tan \kappa \times {\rm tanh}\, \kappa(L-1)= 1\,
 \label{druveq}
 \ee
and yields infinitely many roots again (cf. their graphical
representation in Fig.~\ref{obr1ah}). Hence, the spectrum appears
unbounded from below. This forces us to declare the model manifestly
unphysical.

%> L:=2;plot ({sin(k)/cos(k),cosh(k*(L-1))/sinh(k*(L-1))},k=0..13.73,y=-.71..2);
%L:=1.2;M:=1.33;plot
%({sin(k)/cos(k),cosh(k*(L-1))/sinh(k*(L-1)),
%cosh(k*(M-1))/sinh(k*(M-1))},k=0..11.73,y=-.71..4.2);
%\newpage
\begin{figure}[h]                     %instead of \begin{figure}[t]
\begin{center}                         %instead of \begin{center}
\epsfig{file=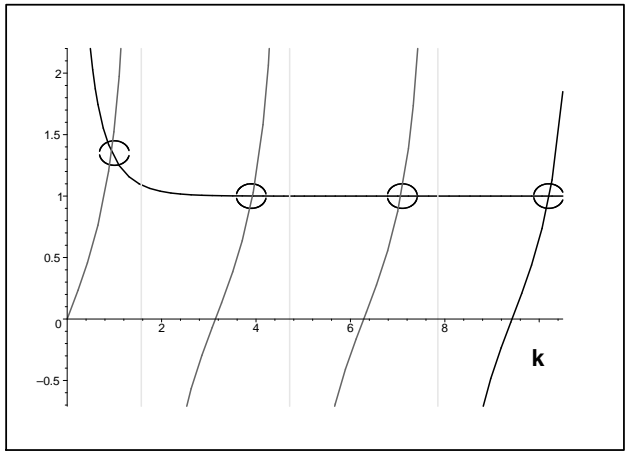,angle=270,width=0.6\textwidth}
\end{center}                         %instead of \end{center}
\vspace{-2mm} \caption{Graphical solution of Eq.~(\ref{druveq}) at
$L=2$ (the small circles mark the intersections of $\tan k$ with
$\coth k(L-1)$).
 \label{obr1ah}}
\end{figure}

The key role of our particular example~(\ref{pushy}) possessing the
discrete bound-state spectrum which is not bounded below should be
seen in its methodical importance. Indeed, although the model is
mathematically correct, its stability will be immediately lost under
the influence of virtually arbitrary perturbation.

Naturally, such a feature of our toy model is generic. Moreover, its
instability under perturbations might have been expected {\em a
priori} since, after the kinetic energy becomes negative, the
perturbed system will certainly have a tendency of plunging,
spontaneously, into the lower and lower energy states. The
corresponding wave functions will acquire a highly oscillatory form.

Qualitatively, the perturbative instability phenomenon will remain
the same for virtually {\em any} external potential $V(x)$. The
exact solvability of our model gives such a sensitivity to
perturbations just a fully explicit form. For our particular
constant and energy-independent choice of the parameter $m_0<0$ one
observes the explicit form of the steady increase of the
localization of the unperturbed wave functions inside the non-empty
interval of $x$. With the growth of parameter $\kappa$ we merely
witness the emergence of an explicit square-well-like part of the
bound-state spectrum which is turned upside down.

%...............

\section{A realistic, energy-dependent model
\label{stabil}}

%The admissibility of negative effective masses

%\subsection{Energy-dependent effective kinetic energy}
%
%
%VVVVVVVVVVVVVVVVVVVVVVVV

In the language of physics, the unperturbed versions of our
simple-minded $m_0<0$ toy model should be re-interpreted as
incomplete. Any realistic realization of similar systems (which, in
principle, radiate, i.e., act as a source of energy) must {\em
necessarily} be reconsidered as coupled to an environment. In other
words, we must re-classify our physical Hilbert space as a mere
subspace ${\cal H}_P$ of a larger physical Hilbert space ${\cal
H}^{full}$.

The lack of the knowledge of the (usually, prohibitively
complicated) full Hamiltonian $H^{full}$ which would characterize
the environment (and which would act in the full Hilbert space
${\cal H}^{full}$) leads to the necessity of acceptance of some
hypotheses. Fortunately, the use of the formalism of the Feshbach's
projection-operator techniques (cf. \cite{Feshbach} for details)
appears extremely efficient in this context.

For our present purposes, in particular, is it sufficient to take
into account that even if the rigorously known spectrum of our
unperturbed, decoupled (or, in the Feshbach's language,
$P-$projected) Hamiltonian of section \ref{instab} is not bounded
from below, {\em any} of its realistic perturbed versions may be
given the Feshbach's semi-explicit form
 \be
 H^{(effective)}=P\,H^{full}\,P+P\,H^{full}\,Q\,
 \frac{1}{E-Q\,H^{full}\,Q}
 \,
 Q\,H^{full}\,P\,
 \label{uSE}
 \ee
where $Q = I-P$. Such an effective Hamiltonian may certainly
represent measurable phenomena but, by assumption, our knowledge of
this operator is restricted and incomplete. Its important merit is
that it is still defined in the accessible, ``small'', $P-$projected
Hilbert space ${\cal H}_P$. Next, the energy $E$ need not be
considered complex in the present setting. The reason is that we may
and shall tacitly assume that the whole spectrum of $H^{full}$
remains discrete and that, for the sake of simplicity, the energy
$E$ in (\ref{uSE}) does not belong to the spectrum of
$Q\,H^{full}\,Q$.

The only conclusion which we can make about the difference between
the schematic, unperturbed negative-mass Hamiltonian (say,
$P\,H^{full}\,P$) and the {\em whole family} of its possible
realistic effective descendants (\ref{uSE}) is that all of the
latter operators must be, {necessarily}, {\em manifestly}
energy-dependent. This fact may be recalled as giving the reasons
why the choice of the effective mass should also be considered
{energy-dependent} in general. {\em Vice versa}, once we admit that
the mass is energy dependent, $m=m(x,E)$, any restriction of
attention just to the positive values of this phenomenological
parameter becomes entirely artificial.

Via our previous, energy-independent example we already understand
that the emergence of the {\em unlimited decrease} of the sequence
of the bound-state energy levels should be attributed to the purely
mathematical role played by the negative constant $m_0<0$ at the
large parameters $E=-\kappa^2$. On this basis we may most simply
stabilize the spectrum by keeping the effective mass positive beyond
certain threshold, $m(x,E)>0$ for $E < E_{thr}$.

In parallel, the physical meaning of the threshold energy cut-off
may vary with the hypotheses concerning the environment. The choice
of its value may be interpreted as a compressed information about
the interplay between the complicated coupling operators
$P\,H^{full}\,Q$  and the Hamiltonian of the environment
$Q\,H^{full}\,Q$. Naturally, even such a weak form of information
about the hidden and/or prohibitively complicated dynamical
mechanisms may still play the role of a phenomenological source of
variability of the effective mass.

Needless to add, the use of the energy-dependent effective operators
finds a broad range of applications in various domains of quantum
physics \cite{Ingrid}. The key feature of their Feshbach's
mathematical origin is that every effective operator (defined in the
$P-$projected subspace) varies with the changes of the energy of the
$(P+Q)-$projected system. In this sense, the unwanted emergence of
mathematical anomalies (like, typically, the unbounded spectrum as
mentioned above) may still be eliminated and attributed to an
inappropriate treatment of the energy-dependent simulation of the
effects of the environment.

%\subsection{The first solvable model and its spectrum}

For the sake of definiteness of our argument let us now proceed by
teaching by example again, replacing the constant parameter $m_0$ in
Eq.~(\ref{pushy}) by its suitable {\em energy-dependent}
generalization. It will be allowed negative at $E > E_{thr}$ for,
say, $E_{thr}=0$. For the sake of simplicity, let us also add the
convenient assumption that $m_0(E) \to 1$ for $E \to -\infty$ and
that $m_0(E) \to -1$ for $E \to +\infty$.

Under these assumptions, we may select, for illustration purposes,
the following, most elementary interpolation ansatz replacing
Eq.~(\ref{pushy}),
 \be
 m(x,E)=
 \left \{
 \begin{array}{cc}
 1,&\ \ \ |x| \in (1,L)\,,\\
 -{\rm tanh} \,(E),\ \ \ & \ \ \ |x| < 1 \,.
 \end{array}
 \right .
 \label{pushybel}
 \ee
Under this choice the value of the mass parameter just simulates the
emergence of an anomaly in the kinetic-energy operator for a
restricted range of the energies. As already mentioned, such an
effective kinetic-energy operator finds its natural hypothetical
origin in the Feshbach's reduction of a realistic or ``complete"
(i.e., formally, $(P+Q)-$projected) Hilbert space (including some
unspecified and formally eliminated ``medium") to its suitable,
explicitly tractable (i.e., $P-$projected {\em alias} model-space)
subspace.

In the resulting amended version
 \be
 -\frac{1}{m(x,E)}
  \frac{d^2}{d{x}^2}\, \Psi(x) = E \,\Psi(x)\,,\ \ \ \ \
 \Psi(-L) = \Psi(L)=0
 \label{SEorexen}
  \ee
of our toy-model Schr\"{o}dinger equation the emergence of
energy-dependence of the mass certainly does not change the
applicability of the matching method of solution. The insertion of
the amended effective mass (\ref{pushybel}) will still enable us to
proceed in an almost complete parallel with the preceding section.
First of all we shall split our equation in its ``outer" and
``inner" part,
 \be
 \begin{array}{cc}
 -  \frac{d^2}{d{x}^2}\, \Psi(x) = E \,\Psi(x)\,\ \ &
  \ \ x \in (-L,-1)
  \,,\\
  - \frac{d^2}{d{x}^2}\, \Psi(x) =  - E\,{\rm tanh} \,(E) \,\Psi(x)\,\ \ &
  \ \ x \in (-1,0)
  \,\
 \ea
 \ee
and restrict our attention, for the sake of brevity, just to the
even-party states again. Next we distinguish between the positive
and negative sign of $E$. In the former case we set $E=k^2$ yielding
the ansatz
 \be
 \begin{array}{cc}
 \Psi(x) = \sin k(x+L)\,,\ \ &
  \ \ x \in (-L,-1)
  \,,\\
   \Psi(x) = D\,{\rm cosh}\,\lambda x\,, \ \ &
  \ \ x \in (-1,0)\,,\ \ \ \ \lambda=\lambda(k)=
  k\,\sqrt{{\rm tanh}\,k^2}>0
  \,,
 \ea
 \ee
while in the latter case we put $E=-\kappa^2$ and get
 \be
 \begin{array}{cc}
 \Psi(x) = {\rm sinh}\, \kappa(x+L)\,,\ \ &
  \ \ x \in (-L,-1)
  \,,\\
   \Psi(x) = F\,{\rm cosh}\,\mu x\,, \ \ &
  \ \ x \in (-1,0)\,,\ \ \ \ \mu=\mu(\kappa)=
  \kappa\,\sqrt{{\rm tanh}\,\kappa^2}>0
  \,.
 \ea
 \ee
The respective matching conditions read
 \be
 \ba
 D\,{\rm cosh}\,\lambda(k)=\sin k(L-1)\,,\\
 -\lambda(k)\,D\,{\rm sinh}\,\lambda(k)=k\,\cos k(L-1)\,,
 \ea
 \ee
 \be
 \ba
 {\rm sinh}\, \kappa (L-1)= F\,{\rm cosh}\, \mu(\kappa)\,,\\
 \kappa\,{\rm cosh}\, \kappa (L-1)= -\mu(\kappa)\,F\,{\rm sinh}\, \mu(\kappa) \,.
 \ea
 \ee
They imply that the eigenvalues may be obtained from the respective
transcendental equations
 \be
   \sqrt{{\rm tanh}\,k^2}\
 {\rm tanh}\,\lambda(k) \, \tan k(L-1)= -1\,,
 \label{prveqpl}
 \ee
 \be
   \sqrt{{\rm tanh}\,\kappa^2}\
 {\rm tanh}\,\mu(\kappa) \,  {\rm  tanh}\,\kappa (L-1)= -1\,.
 \label{prveqmi}
 \ee
We see that the left-hand side of the latter equation is nonnegative
for any real $\kappa$ so that the set of its roots is empty.  The
bound-state energies themselves are all determined by
Eq.~(\ref{prveqpl})  (cf. Fig.~\ref{obr3ag} which may be perceived
as just a slightly deformed analogue of Fig.~\ref{obr1ag}).

\begin{figure}[h]                     %instead of \begin{figure}[t]
\begin{center}                         %instead of \begin{center}
\epsfig{file=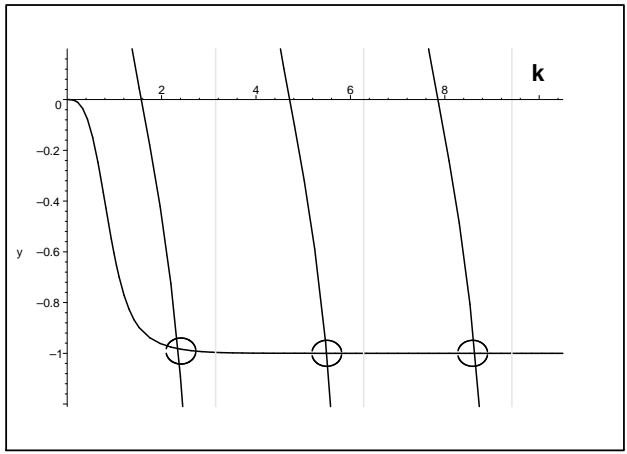,angle=270,width=0.6\textwidth}
\end{center}                         %instead of \end{center}
\vspace{-2mm} \caption{Graphical solution of Eq.~(\ref{prveqpl}) at
$L=2$.
 \label{obr3ag}}
\end{figure}

%lam:= k -> k*sqrt(tanh(k^2));M:=2;plot({
%   -0.991+ sqrt(.1/38-(k-2.41)^2/38),
%  -0.991- sqrt(.1/38-(k-2.41)^2/38),  -1.+ sqrt(.1/38-(k-5.5)^2/38),
%  -1.- sqrt(.1/38-(k-5.5)^2/38), -1.+ sqrt(.1/38-(k-8.59)^2/38),
%  -1.- sqrt(.1/38-(k-8.59)^2/38),
% cos(k*(M-1))/sin(k*(M-1)),-sqrt(tanh(k^2))*sinh(lam(k))/cosh(lam(k))},
%k=0..10.5,y=-1.21..0.1998);

The even-state part of the spectrum proves bounded from below. This
is our main conclusion. Along similar lines, the same conclusion may
be obtained for the odd-parity bound states. We may summarize that
whenever one wants to work with the quantum systems in which the
effective mass can get negative, one is not allowed to neglect the
variability of the effective mass with the energy.

\section{Another exactly solvable model
\label{secondex}}

The introduction of the energy-dependence in the effective mass has
been based on the existence of a hypothetical environment. One would
expect, on this basis, that the ``realistic" $E-$dependence of
$m(x,E)$ would be smooth. In such a case, the study of the related
ordinary differential effective Schr\"{o}dinger equations may
proceed in parallel with the $E-$independent cases. A compact review
of these parallels may be found in Ref.~\cite{energy}. It has been
emphasized there that the overlaps between the wave functions
corresponding to different energies will not vanish in general. In
our present particular model the readers might easily check this
fact using the available explicit formulae for wave functions.

An exhaustive clarification of this apparent paradox may be found in
Ref.~\cite{energy}. Interested readers find there not only the
standard Hilbert-space interpretation of the energy-dependent
Hamiltonians (based on the use of suitably adapted {\em ad hoc}
inner products) but also the explicit realization of the
bi-orthogonality and bi-orthonormality relations between eigenstates
in similar models. One of the most concise reviews of the extension
of this formalism to the case of the general non-Hermitian
observables may be also found in Ref.~\cite{SIGMA}.

We saw, via the schematic illustrative example of preceding section,
that the bound-state spectrum might be very sensitive to the value
of the threshold energy $E_{thr}$. In particular, it may be expected
much more sensitive to this value than to the concrete shape of the
function $m_0(E)$ itself. For this reason, let us now replace the
smooth-function ansatz (\ref{pushybel}) by the simplified but
arbitrarily shifted step-function shape
 \be
 m(x,E)=
 \left \{
 \begin{array}{cc}
 1,&\ \ \ |x| \in (1,L)\,,\\
\left .
 \begin{array}{cc}
 -1, \ \ \ & E \geq E_{thr}\\
 + 1, \ \ \ & E < E_{thr}
 \end{array}
 \right \},
 & \ \ \ |x| < 1 \,
 \end{array}
 \right .
 \label{pushyugl}
 \ee
where the threshold energy is now to be chosen as any finite, real
and, for the sake of definiteness, negative variable $E_{thr} =
-\beta^2$. In this context one must be aware of the fact that for a
generic phenomenological potential given in advance, the similar
discontinuities and abrupt changes of the mass might give rise to
singular contributions to the effective potential. This would
deserve a separate analysis. Interested readers are recommended to
check the existing literature in this respect~\cite{Gangu}.

Certainly, there will be no solutions at the energies $E<-\beta^2$.
For the sake of brevity, we shall again discuss just the even-parity
bound states.  In this case, the  half-infinite interval of the
eligible energies should again be split into the high-energy
half-line with $E = k^2>0$ and the complementary low-energy interval
with $E = -\kappa^2$ where one must merely admit the limited range
of the eligible $\kappa \in (0,\beta)$.

In the former case we may return to Fig.~\ref{obr1ag}, reminding the
readers that the high-energy subset of our present,
``second-example" eigenvalues $E_n=k^2_n$, $n = 1, 2, \ldots$  will
still be defined as roots of the same, unchanged Eq.~(\ref{prveq}).

The similar implicit specification of the second, low-energy subset
of the remaining eigenvalues $E_{1-n}=\kappa^2_n$, $n = 1,2,\ldots,
N$ should again be deduced, {\em mutatis mutandis}, from the
unmodified secular Eq.~(\ref{druveq}). The problem is easily
resolved since a return to Fig.~\ref{obr1ah} reveals that we just
have to replace the original infinite half-axis of $\kappa$ of
section \ref{instab} by its finite subinterval of admissible $\kappa
\in (0,\beta)$.

We may conclude that the new lowest (i.e., ground-state) energy
level will always emerge when there appears a new root of
Eq.~(\ref{druveq}), i.e., the new root $\kappa_N =\beta_{critical}$
of equation
 \be
 \tan \beta_{critical} \times {\rm tanh}\, \beta_{critical}(L-1)=
 1\,.
 \label{druveqbe}
 \ee
In other words, with the increase of the real variable $\beta>0$,
new and new elements of the sequence of the energies
$E_{1-n}=-\kappa_n^2$ computed in section \ref{instab} and sampled
in Fig.~\ref{obr1ah} will be reclassified as the acceptable elements
of the spectrum of the present new model.

Special attention must be paid to the limiting case in which $\kappa
= \beta_{critical}$. It is necessary to keep in mind that also in
such an extreme case the matching condition is satisfied in standard
manner so that the related solution has to be accepted as a valid
new lowest-lying bound state. Perhaps, it is worth adding that as
long as, by our assumption, the effective mass of such a state
remains negative in the whole interval of $x \in (-1,1)$, one should
not be surprised that such a state is more or less fully localized
in this interval. We also see from the explicit formula
(\ref{anonym}) for wave functions that the number of the nodal zeros
of such a wave function, paradoxically, {\em increases} with the
decrease of $E_{thr}=-\beta_{critical}^2$.

\section{Discussion \label{discussion} }

%\section{Towards the more-parametric toy models}

Naturally, our present, methodically motivated restriction of the
number of the free parameters to the necessary minimum might be very
easily relaxed in any future work. One might point out, for example,
that the level-pattern as provided by Fig.~\ref{obr1ah} is, via
secular Eq.~(\ref{druveq}), closely related to the role and
interpretation of the threshold-energy parameter $E_{thr}=-\beta^2$
entering our second illustrative example of paragraph
\ref{secondex}. This means that the {\em same} qualitative picture
will be also provided by the {\em trivially rescaled} models while
the pattern may be changed by an introduction of an additional
parameter.

In such a setting it has been extremely interesting for us to notice
that the replacement of our one-parametric toy model
(\ref{potencal}) + (\ref{pushy}) by its two-parametric
generalization with optional $a>0$ and $b=b(E)>0$ in the effective
mass
 \be
 m(x)=
 \left \{
 \begin{array}{cc}
 1,&\ \ \ |x| \in (a,L)\,,\\
 -1/b^2, \ \ \ & \ \ \ |x| < a\,
 \end{array}
 \right .
 \label{pushytwo}
 \ee
leads to the mere replacement of our previous secular
Eq.~(\ref{druveq}) by the rescaled relation
 \be
 \tan \frac{\kappa a}{b}\times {\rm tanh}\, \kappa(L-a)= b\,.
 \label{druveqsubs}
 \ee
Naturally, no qualitative changes are encountered for the single
rescaling variable $b=a$. Nevertheless, a nontrivial qualitative
change of the pattern will emerge in the ``deep narrow mass well"
regime with small $a$ and $b$ such that also the ratio $a/b:=\nu$
gets small.

In this regime, secular Eq.~(\ref{druveqsubs}) does not possess any
small$-\kappa$ solution so that $\tanh \kappa(L-a) ={\cal O}(1)$.
One reveals that in the way illustrated by Fig.~\ref{obr1ah}, the
sequence of the ``larger" roots $\kappa_2\approx \pi b/a$, $\kappa_3
\approx 2\pi b/a$ ``disappears" to the right infinity in the limit
$\nu=a/b \to 0$. The remaining, single and numerically easily
obtainable leftmost root $\kappa_1$ will remain finite. Its
approximate value will be given by the reduced secular equation
 \be
   \kappa_1 = \frac{b}{\nu}\,{\rm coth}\, \kappa_1 L\,.
 \label{druveqred}
 \ee
Obviously, the right-hand-side expression does not change too
quickly with $\kappa_1$ and it will be, in addition, not too much
larger than the ratio $b/\nu$ itself. Thus, typically, for $b/\nu=1$
one obtains, numerically, the approximate value $\kappa_1\approx
1.2$ of the root.

One just has to add that in such a special $b=\nu$ limit our mass
function (\ref{pushytwo}) with $ -1/b^2\sim 1/a$ may be interpreted
as the Dirac's delta function so that the survival of just the
single bound state in the spectrum might have been, intuitively,
expected.

The most important {\em mathematical} merit of our present choice of
the position-dependent effective masses (which are just piecewise
constant functions of $x$) is that the selection of the rigorously
defined self-adjoint Hamiltonian is traditional and trivial. In the
first two single-parameter toy-models it was provided by the most
elementary matching of the logarithmic derivatives of the wave
functions. This observation has been complemented by the third
example in which we revealed that one could also employ some less
trivial matching prescriptions in certain limiting cases.

Still, no use of the full-fledged self-adjoint-extension theory
\cite{Davies} was needed. Due to the piecewise constant nature of
our effective masses, we did not even have to consider their usual
von Roos' \cite{vanRoos} factorizations $m(x,E)
=m_1(x,E)\,m_2(x,E)\,m_3(x,E)$, nor did we have to select an
appropriate ordering of the individual mass factors and differential
operators $d/dx$ (which, naturally, do not mutually commute in
general).

The resulting simplification of the mathematics and of the necessary
functional analysis did certainly make the physical interpretation
of our models more transparent. This enabled us to emphasize that
the Feshbach-method origin of the concept of the effective mass
opens several new perspectives in the related phenomenology. In our
present letter we managed to demonstrate, first of all, that, and
under which conditions, the effective mass $m=m(x,E)$ need not
necessarily be required a positive function of $x$ and/or $E$.

We might point out that our present detailed description of a few
toy models clarified that there exists an intuitively obvious
connection between the violation of the positivity requirement
$m(x)>0$ and the breakdown of certain traditional theorems. Let us
recall, for example, our observation that in the cases of indefinite
$m(x)$, the ground-state wave function is allowed to possess nodal
zeros so that the traditional Sturm-Liouville oscillation theorems
cease to be valid and must be modified. In parallel, we noticed that
in many cases, the anomalous {\em increase} of the number of the
nodal zeros with the {\em decrease } of the bound state energy is
accompanied by the perceivable {\em localization} of the wave
function inside the interval of negativity of $m(x,E)$.

All of these mathematical observations might find their potential
future physical applications in all of the phenomenological models
where one has to mimic some effects of an (unknown, implicitly
described) external medium by means of the use of the
(energy-dependent) effective operators. In this context we kept in
mind some possible parallels with the relativistic phenomena and
with the well known contrast between the behavior of particles and
antiparticles. Thus, we concentrated our attention to the toy-model
study of the extreme scenario in which the effective kinetic energy
reverses its sign.

We may conclude that the oversimplified nature of our illustrative
examples served our purposes well. In particular, they helped us to
clarify the connection between the {\em decrease} of the value of
the negativity-threshold energy $E_{thr}$ and the {\em increase} of
the number of the ``anomalous", localized and quickly-oscillatory
low-lying bound states. Via our second example admitting a freely
variable $E_{thr}=-\beta^2$, this connection has been even
quantitatively sampled by the elementary-looking restriction of the
pattern given by  Fig.~\ref{obr1ah} to the finite interval of
admissible $\kappa \leq \beta$.

\section*{Acknowledgements}
M. Z. appreciates the support by the GA\v{C}R grant Nr. P203/11/1433
while G. L. acknowledges the support by the OTKA grant Nr. K72357.

\end{document}